\documentclass[12pt]{article}

\usepackage{tikz}
\usepackage{epsfig,latexsym,amsfonts,amsmath,amsthm,amssymb,amsbsy,multirow,slashed,wasysym,textcomp,subfigure,wrapfig,datetime,comment,mathtools,cancel,cite,twistor,mathrsfs}
\usepackage[hidelinks]{hyperref}
\usepackage[font={footnotesize},bf]{caption}

\def\ee{\end{equation}}
\def\be{\begin{equation}}
\def\bea{\begin{eqnarray}}
\def\eea{\end{eqnarray}}
\newcommand{\beq}{\begin{eqnarray}}
\newcommand{\eqq}{\end{eqnarray}}
 \newcommand{\badat}{\begin{alignedat}}
 \newcommand{\eadat}{\end{alignedat}}
\newcommand{\bh}{\bar{h}}
\newcommand{\eal}[1]{\be \begin{aligned} #1 \end{aligned}\end{equation}} 
\newcommand{\eqn}[1]{\be #1 \end{equation}} 
\newcommand{\eqa}[1]{\bea  #1\end{eqnarray}}
\newcommand{\CT}{\mathcal{CT}^2}
\renewcommand{\d}{\mathrm{d}}

\long\def\new#1\endnew{{\bf #1}}		
\long\def\del#1\enddel{}
\def\eps{\epsilon }
\def\del{\partial}

\def\re{\mathrm{e}}

\usepackage{color}

\newcommand{\pink}[1]{\textcolor{\pink}{#1}}

\definecolor{dblue}{rgb}{0.2,0.50,0.80}

\def\A{\mathcal{A}}

\def\vphi{\varphi}

\def\K{\mathbb{K}}
\def\AdS{\text{AdS}}
\newcommand{\br}[1]{\overline{#1}}
\def\veps{\varepsilon}

\def\cL{{\cal L}}
\def\bh{{\bar h}}
\def\bz{{\bar z}}

\def\ba{{\bar a}}
\def\bb{{\bar b}}
\def\bc{{\bar c}}

\def\s{ {\sigma} }

\newcommand{\bs}{\bar{\sigma}}

\numberwithin{equation}{section} 

\begin{document}
\begin{titlepage}
  \thispagestyle{empty}
  \begin{flushright}
    \end{flushright}
  \bigskip
  \begin{center}
	 \vskip2cm
  \baselineskip=13pt {\LARGE \scshape{
  \vspace{0.5em}Celestial Leaf Amplitudes }}

	 \vskip2cm
   \centerline{Walker Melton, Atul Sharma and Andrew Strominger}
 \vskip.5cm
 \noindent{\em Center for the Fundamental Laws of Nature,}
  \vskip.1cm
\noindent{\em  Harvard University,}
{\em Cambridge, MA, USA}
\bigskip
  \vskip1cm
  \end{center}
  \begin{abstract}
  Celestial amplitudes may be decomposed as weighted  integrals of  AdS$_3$-Witten diagrams associated to each leaf of a hyperbolic foliation of spacetime.
  We show, for the  Kleinian three-point MHV amplitude, that each leaf subamplitude is smooth except for the expected light-cone singularities.  Moreover, we find that the full translationally-invariant celestial amplitude  is  simply the residue of  the pole in the leaf amplitude at the point where  the total conformal weights of the gluons equals  three.  This full celestial amplitude vanishes up to  light-cone contact terms, as required  by spacetime translation invariance, and  reduces to the expression previously derived by Mellin transformation of the Parke-Taylor formula. 
  
%
%
  \end{abstract}
%

%
\end{titlepage}
\tableofcontents






\section{Introduction}
\label{sec:intro}

The subleading soft graviton theorem \cite{Cachazo:2014fwa} implies \cite{Kapec:2014opa} that observables of any consistent 4D quantum theory of gravity in asymptotically flat space are invariant  under the action of the local 2D conformal group on the celestial sphere. This brings  the powerful tools of 2D conformal field theory (CFT) to bear on the difficult problem of 4D quantum gravity. In particular the 4D gravitational scattering amplitudes obey the same constraints as those of 2D CFT correlators.

Application of the powerful CFT toolkit  to constrain quantum gravity scattering amplitudes has been hindered by the fact that the latter, although fully conformally invariant, sometimes  take a distributional form which is unfamiliar in  studies of 2D CFT. 
Spacetime translation invariance imposes an infinite number of relations among already-highly-constrained conformally invariant  correlators, and at low points these constraints can be satisfied only by distributional expressions. A number of successful efforts have constructed smooth conformally invariant celestial amplitudes by expanding around  translation noninvariant backgrounds \cite{Costello:2022wso,Fan:2022elem,Casali:2022fro,Gonzo:2022tjm,Stieberger:2023fju,Costello:2022jpg,Stieberger:2022zyk,Melton:2022fsf,Bittleston:2023bzp,Costello:2023hmi,Adamo:2023zeh,Ball:2023ukj}.  

In this paper we show, focusing on the three-gluon MHV amplitude in Klein space\footnote{Similar results apply in Minkowski space with complexified momenta as required for nonzero three-point amplitudes and are related by analytic continuation to those obtained here.}, that the translationally-invariant  distributional expressions have a simple geometric interpretation as sums of generically smooth amplitudes given by AdS$_3/\Z$-Witten diagrams. The derivation given herein is intricate but the basic construction is simple.  Klein space can be foliated by 
two sets of hyperbolic slices in the the timelike and spacelike wedges with $X^2<0$ and $X^2>0$. The leaves of the slices are AdS$_3/\Z$ spacetimes whose boundaries are copies of the celestial torus $\CT$. A conformally invariant \emph{leaf amplitude} on $\CT$ is then given by an AdS$_3/\mathbb{Z}$-Witten diagram for each leaf. These are not constrained by translation invariance as an individual leaf preserves only Lorentz/conformal symmetry. They take the familiar 2D CFT form\footnote{The general structure of a 2D CFT correlator on a Lorentzian torus is discussed  in \cite{Melton:2023hiq}.} and are smooth up to light-cone singularities. To get the full celestial amplitude, we must integrate over all the leaves. We verify, as required by translation invariance, that the full amplitude vanishes for generic scattering angles due to cancellations between the two wedges.

Care must be exercised when two points lie on a common light cone in $\CT$. In this case we show that the cancellation between the two wedges is incomplete. A  contact interaction arises from three-point generalizations of the basic identity
\be    {\im \over \s+\im\eps}-{\im \over \s-\im\eps}=2\pi\, \delta(\s),\ee
where $\eps$ is a regulator for a light cone singularity in a null coordinate $\s$. Moreover we find the strikingly  simple result that the full amplitude is proportional to the pole appearing in the leaf amplitudes when the net sum of the conformal weights of the gluons equals three. This expression agrees in full detail with the PSS expression \cite{Pasterski:2017ylz} for three-gluon MHV scattering obtained via Mellin transforms of the Parke-Taylor formula. Our main result for the $\im\eps$-regulated 3-gluon celestial amplitude is equation \eqref{A3broken}. 

Leaf amplitudes are simpler than the full celestial amplitudes which they are used to construct. Moreover, they have a natural holographic interpretation  provided by the $\AdS_3$/CFT$_2$ correspondence. This suggests that leaf amplitudes may provide useful building blocks for holography 
in flat space. Also of interest in this regard are the \emph{half amplitudes} obtained by integrating the leaf amplitudes over one of the two wedges.\footnote{In either case translation invariance would emerge  at a later stage \cite{cmsw}. }

In order to be specific this paper considers only the MHV 3-point amplitude, but we anticipate  general constructions of celestial from leaf amplitudes at all orders for any number of particles. We have not considered this in detail.

The rest of this paper is organized as follows. Section \ref{sec:prelims} briefly reviews Klein space and celestial amplitudes, collects useful formulae and sets our conventions. Section \ref{sec:celestial} defines leaf amplitudes, relates them to AdS$_3$-Witten diagrams and expresses the full celestial amplitudes as weighted integrals over their leaves.  The gluon leaf amplitude turns out to be a kinematic factor times a scalar contact Witten diagram. This scalar contact diagram is evaluated in section \ref{sec:leaf}, with care to keep track of the necessary $\im\eps$ prescriptions in Lorentzian signature. In section \ref{sec:pss} we show that only the simple pole in the leaf amplitude at the point where the gluon conformal weights sum to 3 contributes to the full celestial amplitude. Moreover we find that the coefficient of this pole contains delta functions in the light cone coordinates and matches perfectly  the known PSS Mellin transform expression,  including the indicator functions separating  causal regions.


\section{Preliminaries}
\label{sec:prelims}

\subsection{Klein space kinematics}

Let us start by reviewing the geometry of Klein space $\K = \R^{2,2}$ and its foliation by Lorentzian $\AdS_3/\Z$ slices or `leaves'. Its conformal boundary, where the dual CCFT resides, is foliated by celestial tori. Points on the celestial torus parametrize Kleinian null momenta. For further discussion, see \cite{Atanasov:2021oyu, Mason:2005qu}.

\paragraph{Hyperbolic foliation of Klein space.} The metric of Klein space with coordinates $X^\mu$ reads
\be\label{Kmet}
\d s^2 = -\,(\d X^0)^2 - (\d X^1)^2 + (\d X^2)^2 + (\d X^3)^2\,.
\ee
Upon excising the light cone of the origin, Klein space splits into a timelike wedge $W^T$ on which $X^2<0$, and a spacelike wedge $W^S$ on which $X^2>0$. Depending on the region, we write
\be
\begin{split}
    W^T\;:\quad X^\mu = \tau\hat x_+^\mu\,,\qquad\hat x_+^2=-1\,,\\
    W^S\;:\quad X^\mu = \tau\hat x_-^\mu\,,\qquad\hat x_-^2=+1\,,
\end{split}
\ee
where $\tau\in(0,\infty)$ is the magnitude of $X^\mu$. In either region, slices of constant $\tau$ are diffeomorphic to $\AdS_3/\Z$, i.e., Lorentzian anti-de Sitter space with periodically identified time. Taken together, they constitute a hyperbolic foliation of Klein space, mirroring  the hyperbolic foliation of Minkowski space used by de Boer and Solodukhin \cite{deBoer:2003vf}.
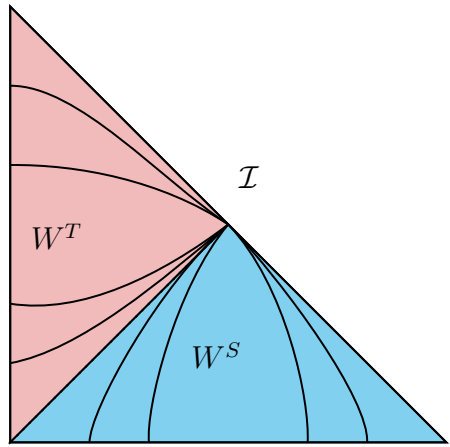
\begin{figure}[h!]
\begin{center}

\tikzset{every picture/.style={line width=0.75pt}} 

\begin{tikzpicture}[x=0.75pt,y=0.75pt,yscale=-1,xscale=1]
\draw  [fill={rgb, 255:red, 241; green, 187; blue, 187 }  ,fill opacity=0.25 ] (210,140) -- (100,250) -- (100,130) -- (100,30) -- (100,30) -- cycle ;
\draw  [fill={rgb, 255:red, 129; green, 208; blue, 240 }  ,fill opacity=0.25 ] (210,140) -- (320,250) -- (100,250) -- (210,140) -- (210,140) -- cycle ;
\draw   (100,30) -- (320,250) -- (100,250) -- cycle ;
\draw  [dash pattern={on 0.84pt off 2.51pt}]  (100,250) -- (210,140) ;
\draw    (100,210) .. controls (135.14,203.65) and (177.14,167.65) .. (210,140) ;
\draw    (100,180) .. controls (136.14,185.65) and (176.14,165.65) .. (210,140) ;
\draw    (100,110) .. controls (134.14,109.65) and (176.14,117.65) .. (210,140) ;
\draw    (100,70) .. controls (134.14,69.65) and (176.14,117.65) .. (210,140) ;
\draw    (140,250) .. controls (140.14,233.65) and (175.14,170.65) .. (210,140) ;
\draw    (170,250) .. controls (168.14,233.65) and (183.14,171.65) .. (210,140) ;
\draw    (250,250) .. controls (251.14,228.65) and (239.14,170.65) .. (210,140) ;
\draw    (280,250) .. controls (281.14,228.65) and (239.14,170.65) .. (210,140) ;

\draw (109,138) node [anchor=north west][inner sep=0.75pt]    {$W^{T}$};
\draw (190,198) node [anchor=north west][inner sep=0.75pt]    {$W^{S}$};
\draw (214,109) node [anchor=north west][inner sep=0.75pt]    {$\mathcal{I}$};

\end{tikzpicture}
\caption{A toric Penrose diagram depicting the hyperbolic foliation of Klein space in the timelike ($W^T$) and spacelike ($W^S$) regions.}
\end{center}
\end{figure}

The timelike leaves $\hat x^2_+=-1$ which foliate the timelike wedge $W^T$ can be parametrized in terms of standard global coordinates:
\be
\hat x_+^0+\im\hat x_+^1 = \re^{\im\psi}\cosh\rho\,,\qquad\hat x_+^2+\im\hat x_+^3 = \re^{\im\vphi}\sinh\rho\,.
\ee
Here, $\rho\in(0,\infty)$ is the radial coordinate on $\AdS_3/\Z$, whereas $\psi\sim\psi+2\pi$ and $\vphi\sim\vphi+2\pi$ are periodic coordinates along timelike and spacelike cycles respectively. The metric is
\be \label{met} \d s^2=-\,\d\tau^2 +\tau^2\big(\!-\cosh^2\!\rho\, \d\psi^2+\sinh^2\!\rho \,\d\vphi^2+\d\rho^2\big)\,,\ee
where the bracketed part is the unit metric on AdS$_3/\Z$. 

The spacelike wedge $W^S$ ($\hat x_-^2=+1$) can be parametrized by analogous coordinates, but with the roles of spacelike and timelike cycles exchanged:
\be
\hat x_-^0+\im\hat x_-^1 = \re^{\im\psi}\sinh\rho\,,\qquad\hat x_-^2+\im\hat x_-^3 = \re^{\im\vphi}\cosh\rho\,.
\ee
In this region, the metric becomes
\be \d s^2=\d\tau^2 -\tau^2\big(\!-\cosh^2\!\rho\, \d\vphi^2+\sinh^2\!\rho \,\d\psi^2+\d\rho^2\big)\,.\ee

The $\rho\to\infty$ conformal boundary of each AdS$_3/\Z$ slice is a Lorentzian torus $\CT=S^1\times S^1$ referred to as the  \emph{celestial torus}. In the coordinates $\psi,\vphi$ on $\CT$, it is equipped with the induced conformal metric $\d s^2=-\,\d\psi^2+\d\vphi^2$. It also admits null coordinates
\be
\s = \frac{\psi+\vphi}{2}\,,\qquad\bs = \frac{\psi-\vphi}{2}
\ee
in terms of which the conformal metric becomes $\d s^2=-4\,\d\s\,\d\bs$. They are subject to the coupled periodic identifications
\be
(\s,\bs)\sim(\s+\pi(m+n),\bs+\pi(m-n))\,,\qquad m,n\in\Z\,.
\ee
In the study of Lorentzian (celestial) CFT, one conventionally works with planar coordinates $z,\bz\in\R$ defined by
\be
z = \tan\s\,,\qquad\bz = \tan\bs
\ee
on which the bulk isometry group $\SL(2,\R)\times\br\SL(2,\R)$ acts by real and independent M\"obius transformations. But these only cover half the celestial torus as they do not distinguish between antipodal points $(\psi,\vphi)$ and $(\psi+\pi,\vphi+\pi)$. Consequently, we will find it useful to work directly in terms of the global coordinates $\s,\bs$.

Geometrically, $z,\bz$ are local coordinates on a Minkowski diamond $\R^{1,1}$, as the 2D metric reduces to $\d s^2 = -\d z\,\d\bz $ up to conformal rescalings. The open set $\CT-\{z\bz=0\}$ of the celestial torus is covered by two such diamonds. On this set, if $\bs$ is brought to its fundamental domain $0<\bs<\pi$, then the two diamonds are cut out by the coordinate regions $0<\sigma<\pi$ and $\pi<\s<2\pi$. Since the celestial torus has only one connected component, the distinction between the two diamonds is purely a choice of coordinate patches and evaporates in  global coordinates.

\paragraph{Massless kinematics.} Momenta of massless particles in $(2,2)$ signature are parametrized in planar coordinates as
\be\label{ppar}
p_i^\mu = \veps_i\omega_iq_i^\mu\,.
\ee
Here $i$ is a particle label, $\omega_i\in(0,\infty)$ denotes the magnitude of the frequency, $\veps_i\in\{\pm1\}$ denotes the sign of the frequency, and $q_i^\mu$ are null vectors
\be
q_i^\mu = \big(1-z_i\bz_i,z_i+\bz_i,1+z_i\bz_i,z_i-\bz_i\big)
\ee
that are normalized to satisfy $q_i^0+q_i^2=2$. These are null with respect to the $(--+\,+)$ signature flat metric \eqref{Kmet}. 
$(z_i,\bz_i)$ describe points on the celestial torus where the massless particles exit. The choice of coordinates $z,\bz$ divides the torus into two diamonds, and the signs $\veps_i$ denote the choice of the diamond. This is equivalent to a choice of in $vs.$ out in  Minkowski space, and we occasionally continue to use this language for sake of simplicity.

In this work,  instead of planar we primarily use global coordinates on the celestial torus. Substituting
\be
z_i = \tan\s_i\,,\qquad\bz_i=\tan\bs_i\,,\qquad\s_i = \frac{\psi_i+\vphi_i}{2}\,,\qquad\bs_i =\frac{\psi_i-\vphi_i}{2}\,,
\ee
we find the decomposition 
\be\label{pglob}
p_i^\mu = \frac{\omega_i\hat p_i^\mu}{|\cos\s_i\cos\bs_i|}\,,
\ee
where $\hat p_i^\mu$ are null vectors parametrized by points $(\psi_i,\vphi_i)\in\CT$:
\be
\label{eq:masslessparam}
\hat p_i^0+\im\hat p_i^1 = \re^{\im\psi_i}\,,\qquad\hat p_i^2+\im\hat p_i^3 = \re^{\im\vphi_i}\,. 
\ee
In particular, we identify $\veps_i = (p_i^0+p_i^2)/2\omega_i = \sgn(\cos\s_i\cos\bs_i)$.

The inner products of such a null vector $\hat p_i^\mu$ with the unit timelike or spacelike vectors $\hat x_\pm^\mu$ are found to be
\begin{align}
    \hat p_i\cdot\hat x_+ &= \cos(\vphi-\vphi_i)\sinh\rho - \cos(\psi-\psi_i)\cosh\rho\,,\label{pi.x+}\\
    \hat p_i\cdot\hat x_- &= \cos(\vphi-\vphi_i)\cosh\rho - \cos(\psi-\psi_i)\sinh\rho\,.\label{pi.x-}
\end{align}
 Similarly, the inner product of two null vectors $\hat p_i^\mu,\hat p_j^\mu$ is given by
\be\label{pipj}
    \hat p_i\cdot\hat p_j = \cos\vphi_{ij}-\cos\psi_{ij}= 2\sin\s_{ij}\sin\bs_{ij}\,,
\ee
where $\vphi_{ij}=\vphi_i-\vphi_j$, $\s_{ij}=\s_i-\s_j$, etc. We will also find it useful to introduce the following abbreviations for the torus separations:
\be\label{sdef}
s_{ij} \vcentcolon = \sin\s_{ij}\,,\qquad \bar s_{ij} \vcentcolon= \sin\bs_{ij}\,.
\ee
These become the natural variables for writing CFT correlation functions on Lorentzian tori \cite{Melton:2023hiq}.


\subsection{Celestial amplitudes in global coordinates}


Here we give our conventions for celestial amplitudes. Let $A(p_i,J_i)$ denote a momentum space scattering amplitude for massless particles carrying momenta $p_i$ and helicities $J_i$, where $i=1,\dots,n$. In particular, this contains the momentum conserving delta function $\delta^4(p_1+\cdots+p_n)$. The corresponding celestial amplitude is given  by the Mellin transform, 
\be\label{camp}
\cA(\veps_i,\Delta_i,J_i,z_i,\bz_i) = \prod_{j=1}^n\int_0^\infty\d\omega_j\,\omega_j^{\Delta_j-1}\re^{-\eps\,\omega_j}\,A(\veps_i\omega_iq_i,J_i)\,,
\ee
where $\eps>0$ is a small regulator. It depends on the celestial torus positions $z_i,\bz_i$, on the signs $\veps_i$, as well as on a choice of conformal weights $\Delta_i\in\C$. Under 2D conformal transformations, it transforms as a 2D CFT correlator   \cite{Pasterski:2016qvg}\be
\cA\bigg(\veps_i,\Delta_i,J_i,\frac{az_i+b}{cz_i+d},\frac{\bar a\bz_i+\bar b}{\bar c\bz_i+\bar d}\bigg) = \prod_{j=1}^n|cz_j+d|^{2h_j}|\bar{c}\bz_j+\bar{d}|^{2\bh_j}\,\cA(\veps_i,\Delta_i,J_i,z_i,\bz_i)\,,
\ee
where 
\be
(h_i,\bh_i) = \bigg(\frac{\Delta_i+J_i}{2},\frac{\Delta_i-J_i}{2}\bigg).
\ee
 are the left/right conformal weights.

In Lorentzian signature, a choice of complete basis for the space of external states is provided by conformal primary states with weights ranging over the principal series \cite{Pasterski:2017kqt}: $\Delta_i\in1+\im\,\R$. Whenever needed, our expressions will be understood as being defined for the same choice of weights in $(2,2)$ signature as well.

To rewrite celestial amplitudes in global coordinates, we substitute the global parametrization \eqref{pglob} into the definition \eqref{camp} and rescale $\omega_i\mapsto\omega_i|\cos\s_i\cos\bs_i|$ for each $i$. This produces
\be\label{ltog}
\cA(\veps_i,\Delta_i,J_i,z_i,\bz_i) = \prod_{j=1}^n|\cos\s_j|^{2h_j}|\cos\bs_j|^{2\bh_j}\,\cA(\Delta_i,J_i,\s_i,\bs_i)\,,
\ee
where we have defined the torus uplifts of massless celestial amplitudes:
\begin{equation}\label{campg}
    \cA(\Delta_i,J_i,\s_i,\bs_i) = \prod_{j=1}^n\int_0^\infty\d\omega_j\,\omega_j^{\Delta_j-1}\re^{-\eps\,\omega_j}\,A(\omega_i\hat p_i,J_i)\,.
\end{equation}
Equivalently, this is just the celestial amplitude for scattering particles carrying null momenta $p_i=\omega_i\hat p_i$. 

If one has already computed the celestial amplitude $\cA(\veps_i,\Delta_i,J_i,z_i,\bz_i)$, one can also derive its torus uplift $\cA(\Delta_i,J_i,\s_i,\bs_i)$ by simply plugging in
\be
z_{ij} = \frac{\sin\s_{ij}}{\cos\s_i\cos\s_j}\,,\qquad \bz_{ij} = \frac{\sin\bs_{ij}}{\cos\bs_i\cos\bs_j}
\ee
and dividing out the Jacobian factor $\prod_j|\cos\s_j|^{2h_j}|\cos\bs_j|^{2\bh_j}$. Crucially, as we are working in global coordinates, the result no longer depends on any ingoing or outgoing labels $\veps_i$ because the parametrization in \eqref{eq:masslessparam} covers the entire celestial torus.

\paragraph{3-point PSS amplitude.} Our primary object of interest is the celestial amplitude for three gluons in the maximally helicity violating (MHV) sector. We take this to be the amplitude where gluons $1,2$ have helicity $J_1=J_2=-1$ and gluon $3$ has helicity $J_3=+1$. To compute this, one starts with the Parke Taylor formula for the associated momentum space amplitude:
\be
A(1^-2^-3^+) = \frac{\la 12\ra^3}{\la23\ra\la31\ra}\;\delta^4(p_1+p_2+p_3)\,,
\ee
where we are only considering the color-stripped amplitude. Here
\be
\la ij\ra = \sqrt{\omega_i\omega_j}z_{ij}\,,\qquad z_{ij}=z_i-z_j\,,
\ee
are the so-called spinor-helicity brackets that are commonly employed in computations of massless scattering \cite{Elvang:2017}. Also, $A(1^{J_1}2^{J_2}3^{J_3}\cdots)$ is short for $A(p_1,J_1;p_2,J_2;p_3,J_3;\cdots)$, and we will use a similar abbreviation when writing celestial amplitudes.

On plugging in the parametrization \eqref{ppar} and performing the Mellin transforms, one finds the 3-point PSS amplitude \cite{Pasterski:2017ylz,Pate:2019mfs}
\be\label{A3}
\cA(1^-2^-3^+) = \frac{\pi}{2}\,\delta(\beta)\,\frac{\sgn(z_{12}z_{23}z_{31})}{|z_{12}|^a|z_{23}|^b|z_{31}|^c}\,\delta(\bz_{13})\,\delta(\bz_{23})\,\Theta\bigg(\frac{\veps_3z_{23}}{\veps_1z_{12}}\bigg)\,\Theta\bigg(\frac{\veps_3z_{31}}{\veps_2z_{12}}\bigg)\,,
\ee
where $\Theta(x)$ is the Heaviside step function, and $|z_{ij}|$ are real absolute values of the chiral separations $z_{ij}$.\footnote{They are not to be confused with $(z_{ij}\bz_{ij})^\frac{1}{2}$.} Here, \be \beta = 2(\bh_1+\bh_2+\bh_3-2),\ee and we have introduced the convenient notation 
\be
a = h_1+h_2-h_3\,,\qquad b = h_2+h_3-h_1\,,\qquad c = h_3+h_1-h_2\,,
\ee
along with their antichiral counterparts
\be\label{bardef}
\ba = \bh_1+\bh_2-\bh_3\,,\qquad \bb = \bh_2+\bh_3-\bh_1\,,\qquad \bc = \bh_3+\bh_1-\bh_2\,.
\ee
On the support of $\delta(\beta)$, these exponents obey $a+b+c=1$ and $\ba+\bb+\bc=2$.

In global coordinates, the PSS amplitude takes the similar form
\begin{equation}
\label{eq:pssglobal}
\begin{split}
    \A(1^-2^-3^+) = \frac{\pi}{2}\,\delta(\beta)&\,\frac{\sgn(\sin\s_{12}\sin\s_{23}\sin\s_{31})}{|\sin\s_{12}|^a|\sin\s_{23}|^b|\sin\s_{31}|^c}\,\delta(\sin\bs_{13})\,\delta(\sin\bs_{23}) \\
&\times\Theta\left(\frac{\cos\bs_{23}}{\cos\bs_{12}}\frac{\sin\s_{23}}{\sin\s_{12}}\right)\Theta\left(\frac{\cos\bs_{31}}{\cos\bs_{12}}\frac{\sin\s_{31}}{\sin\s_{12}}\right)\,.
\end{split}
\end{equation}
Because we are using global coordinates, the ingoing/outgoing labels $\veps_i$ have been eliminated using the relations $\veps_i=\sgn(\cos\s_i\cos\bs_i)$. We have also used $\sgn(\cos\bs_i\cos\bs_j)=\sgn(\cos\bs_{ij})$ which is valid on the support of $\delta(\sin\bs_{13})\,\delta(\sin\bs_{23})$. The step functions ensure that there is no way to divide the celestial torus into an ingoing and outgoing diamond such that all three particles are ingoing or outgoing when continued to Lorentzian signature. 

It is interesting and nontrivial to show that \eqref{eq:pssglobal} is single-valued and conformally covariant on $\CT$ 
\cite{Pasterski:2017ylz,Melton:2023hiq}, as will be reconfirmed in the derivation herein. In fact these considerations largely fix \eqref{eq:pssglobal}. 


\section{Leaf  amplitudes}
\label{sec:celestial}

Celestial amplitudes are often constructed  as Mellin transforms of momentum space amplitudes. Alternately,  evaluating celestial amplitudes directly in position space usefully presents them as weighted integrals of  Witten diagrams on the leaves of a hyperbolic foliation of spacetime  \cite{Pasterski:2016qvg,Casali:2022fro,Iacobacci:2022yjo}, or `leaf amplitudes'.  
In this section we define the leaf amplitude arising in Kleinian MHV gluon scattering, but we expect the construction to generalize. 

The color-stripped momentum space Parke-Taylor amplitude is
\be
A(1^-2^-3^+\cdots n^+) = \frac{\la12\ra^3}{\la23\ra\la34\ra\cdots\la n1\ra}\;\delta^4(P)\,,
\ee
where $P^\mu$ denotes the total momentum
\be
P^\mu = \sum_{i=1}^n p_i^\mu\,.
\ee
We wish to express the associated celestial amplitude directly as an integral over conformal primary wavefunctions without reference to momentum space. 

We start by invoking the Fourier representation of the momentum-conserving delta function,
\begin{equation}
    \delta^4(P) = \int\frac{\d^4X}{(2\pi)^4}\; \re^{\im P\cdot X}\,.
\end{equation}
We then perform  the Mellin transform \eqref{camp} before the spacetime integration. On using $z_{ij}=\sin\s_{ij}/\cos\s_i\cos\s_j$, the expression for angle brackets in global coordinates is found by stripping off the $\cos\s_i\cos\s_j$ factors\footnote{These always get absorbed in the Jacobian of the conformal map $z_i=\tan\s_i$, $\bz_i=\tan\bs_i$ displayed on the right hand side of \eqref{ltog}.}
\be
\la ij\ra = \sqrt{\omega_i\omega_j}\sin\s_{ij}\,.
\ee
Using the notation \eqref{sdef}, the MHV celestial amplitude \eqref{camp} then takes the desired form 
\begin{equation}
\label{eq:posamp}
    \begin{split}
    \A(1^-2^-3^+\cdots n^+) &=  \prod_{j=1}^n\int_0^\infty\d\omega_j\,\omega_j^{\Delta_j-1}\, \frac{\langle 12\rangle^3}{\langle 23 \rangle\cdots \langle n1\rangle} \int \frac{\d^4X}{(2\pi)^4}\;\re^{\sum_j\omega_j(\im \hat p_j\cdot X-\eps)} \\
    &= \frac{s^3_{12}}{s_{23}\cdots s_{n1}}\int \frac{\d^4X}{(2\pi)^4}\;\prod_{i=1}^n\Phi_{2\bh_i}(X,\hat p_i)\,,
    \end{split}
\end{equation}
where the scalar conformal primary wavefunctions are
\begin{equation}\label{fd}
    \Phi_\Delta(X,\hat{p}) = \int_0^\infty \d\omega\,\omega^{\Delta-1}\,\re^{\im\omega\hat{p}\cdot X - \eps\,\omega} = \frac{\Gamma(\Delta)}{(-\im\hat{p}\cdot X + \eps)^\Delta}\,.
\end{equation}
The Mellin integral here is defined with a small, positive regulator $\eps$. In hyperbolic coordinates, these wavefunctions turn into embedding space expressions for bulk-to-boundary propagators on Lorentzian AdS$_3/\Z$.

The expression \eqref{eq:posamp} transforms as a correlation function of $n$ conformal primaries of weights
\begin{equation}
    \begin{split}
        h_1 &= \frac{\Delta_1-1}{2}\,,\quad \bar{h}_1 = \frac{\Delta_1+1}{2}\,, \\
        h_2 &= \frac{\Delta_2-1}{2}\,,\quad \bar{h}_2 = \frac{\Delta_2+1}{2}\,, \\
        h_j &= \frac{\Delta_j+1}{2}\,,\quad \bar{h}_j = \frac{\Delta_j-1}{2}\,,\quad j = 3,\ldots, n\,.
    \end{split}
\end{equation}
We then have
\be
\beta = \sum_{j=1}^n(\Delta_j-1) = -4+2\sum_{j=1}^n\bar{h}_j\,.
\ee
The next step is to break the integral over Klein space into two integrals, one over the timelike wedge  $W^T$ ($X^2<0$) and another over the spacelike wedge $W^S$ ($X^2>0$). In either region, the integral in \eqref{eq:posamp} can be factorized into an integral over $\tau = \sqrt{|X^2|}\in(0,\infty)$, and an integral over a unit AdS$_3/\mathbb{Z}$ slice. This yields
\begin{multline}
    \A(1^-2^-3^+\cdots n^+) = \frac{s^3_{12}}{s_{23}\cdots s_{n1}}\int_0^\infty\d\tau\,\tau^3\\
    \times\bigg\{\int_{\hat x_+^2=-1}\d^3\hat x_+\prod_{i=1}^n\Phi_{2\bh_i}(\tau\hat x_+,\hat p_i) + \int_{\hat x_-^2=+1}\d^3\hat x_-\prod_{i=1}^n\Phi_{2\bh_i}(\tau\hat x_-,\hat p_i)\bigg\}\,,
\end{multline}
where $\d^3\hat x_\pm$ are the volume forms on $\AdS_3/\Z$ (see \eqref{volpm} below). Additionally, a simple change of variables shows that the integral over the spacelike region of Klein space can be obtained from the integral over the timelike region by sending $\bs_i \mapsto -\bs_i$. 

Since $\tau>0$, in the limit $\eps\to0$ we can use $\Phi_\Delta(\tau\hat x_\pm,\hat p_i) = \tau^{-\Delta}\Phi_\Delta(\hat x_\pm,\hat p_i)$ to rewrite this as
\begin{align}
    \A(1^-2^-3^+\cdots n^+) &= \frac{\delta(\beta)}{8\pi^3}\,\big(\cL(\s_i,\bs_i) + \cL(\s_i,-\bs_i)\big). \label{leafy0}\\
    \cL(\s_i,\bs_i)&=\frac{s^3_{12}}{s_{23}\cdots s_{n1}}\;\cC(\s_i,\bs_i) \label{leafy} \\
    \cC &= \int_{\AdS_3/\Z} \d^3\hat{x}_{+}\prod_{i=1}^n\Phi_{2\bar{h}_i}(\hat{x}_+,\hat{p}_i)\label{Cpm}
\end{align}
$\cL$ here  is the MHV $n$-point \emph{leaf amplitude}, which is proportional to the leaf amplitude $\cC$ for $n$  massless scalars. 
The $\tau$ integral in \eqref{leafy0} has been performed using the identity
\be
\int_0^\infty\d\tau\,\tau^{-\beta-1} = 2\pi\,\delta(\beta)\,.
\ee
Strictly speaking, this holds when $\text{Re}\,\beta=0$, and $\delta(\beta)$ is to be interpreted as $\delta(\text{Im}\,\beta)$. But it can be analytically continued to other $\beta$ as explained in \cite{Donnay:2020guq}. 

As a result, the integral over $\tau$ reproduces the $\delta(\beta)$ expected from bulk scale invariance, while the integrals over the unit $\AdS_3/\Z$ slices $\hat x^2_\pm=\mp1$ generate leaf amplitudes $\cL(\s_i,\pm \bs_i)$. We see here that the conformal weights for leaf amplitudes are unconstrained: the celestial amplitude constraint on their total sum (for massless particles) comes from the $\tau$ integral over leaves.  Also of interest are the nontranslationally invariant half amplitudes $\cal H$, obtained by integrating over the leaves in only the timelike wedge $W^T$. These include the  constraint on the weights and are simply related to the full amplitudes by  \be  \A(1^-2^-3^+\cdots n^+) = {\cal H}(\s_i,\bs_i) + {\cal H}(\s_i,-\bs_i).\ee

We will now evaluate these leaf amplitudes for 3-gluon MHV scattering and show that \eqref{leafy0} reproduces the standard gluon 3-point celestial amplitude \eqref{A3}, even though the leaf amplitudes themselves are non-distributional. 


\section{3-point scalar leaf amplitude}
\label{sec:leaf}

The leaf amplitude \eqref{Cpm} is a  contact Witten diagram for massless scalars propagating on Lorentzian $\AdS_3/\Z$. At three points we have 
\be
\cC = \int_{\AdS_3/\Z}\d^3\hat x_+\prod_{i=1}^3\frac{\Gamma(2\bh_i)}{(\eps-\im\hat p_i\cdot\hat x_+)^{2\bh_i}}\,.
\ee
Here $\hat x_+^2=-1$ as before, and $\eps$ is a small, positive regulator that picks a choice of branch for each bulk-to-boundary propagator. To be precise, for $\Delta\in\C$ and $x\in\R-0$, quantities like $(x\pm\im\eps)^\Delta$ are defined by\footnote{See section 3.6 of \cite{Gelfand1} for a useful review.}
\be\label{iedef}
(x\pm\im\eps)^\Delta = \begin{cases}
    x^\Delta\qquad &x>0\\
    \re^{\pm\im\pi\Delta}(-x)^\Delta\qquad &x<0
\end{cases}
\ee
understood in the limit $\eps\to0^+$. The measure of integration is
\be\label{volpm}
\d^3\hat x_+ = \sinh\rho\cosh\rho\,\d\rho\,\d\psi\,\d\vphi\,.
\ee
We will use  $\psi,\vphi$ as integration variables instead of $\s,\bs$ as this will aid us in separation of integrals. 
The constraint that $\beta = 0$ arises from $\tau$ integration and does not restrict the leaf amplitudes of this section. 

\paragraph{Hyperbolic integrals.} To evaluate the AdS$_3/\Z$ integrals,  we reinstate the Mellin representations of the bulk-to-boundary propagators:
\be
\cC = \int_{\AdS_3/\Z}\d^3\hat x_+\prod_{i=1}^3\int_0^\infty\d\omega_i\,\omega_i^{2\bh_i-1}\,\re^{\im\omega_i\hat p_i\cdot\hat x_+ - \eps\,\omega_i}\,.
\ee
Using \eqref{eq:masslessparam}, this becomes
\begin{multline}\label{C0}
    \cC = \int_0^\infty\d\rho\,\sinh\rho\cosh\rho\int_0^{2\pi}\d\psi\int_0^{2\pi}\d\vphi\\
    \times\prod_{i=1}^3\int_0^\infty\d\omega_i\,\omega_i^{2\bh_i-1}\exp\Big\{\im\omega_i\cos(\vphi-\vphi_i)\sinh\rho-\im\omega_i\cos(\psi-\psi_i)\cosh\rho-\eps\,\omega_i\Big\}\,.
\end{multline}
We now substitute $\cos(\vphi-\vphi_i) = \cos\vphi\cos\vphi_i+\sin\vphi\sin\vphi_i$ and similarly expand out $\cos(\psi-\psi_i)$. Using the  Bessel function representation
\be
\int_0^{2\pi}\d\vphi\;\re^{\im x\cos\vphi+\im y\sin\vphi} = 2\pi\,J_0(\sqrt{x^2+y^2})\,,
\ee
we can perform the angular $\psi,\vphi$ integrals:
\be\label{C1}
    \cC = 4\pi^2\prod_{i=1}^3\int_0^\infty\d\omega_i\,\omega_i^{2\bh_i-1}\,\re^{-\eps\,\omega_i}\int_0^\infty\d y\,y\, J_0(\Phi y)\,J_0(\Psi\sqrt{1+y^2})\,.
\ee
Here we have substituted $y=\sinh\rho$ and abbreviated
\be\
\begin{split}
    \Phi &\equiv \sqrt{\textstyle\sum_j\omega_j^2+2\textstyle\sum_{j<k}\omega_j\omega_k\cos\vphi_{jk}}\,,\\
    \Psi &\equiv \sqrt{\textstyle\sum_j\omega_j^2+2\textstyle\sum_{j<k}\omega_j\omega_k\cos\psi_{jk}}\,.
\end{split}
\ee
These have been expressed in terms of $\vphi_{jk}=\vphi_j-\vphi_k$ and $\psi_{jk}=\psi_j-\psi_k$.

The integral over $y$ can be performed by using the Schl\"afli integral representation of the Bessel function,\footnote{This technique for performing the $y$ integral is taken from section 13.47 of \cite{Watson:1922}.}
\be
J_\nu(z) = \frac{1}{2\pi\im}\bigg(\frac{z}{2}\bigg)^\nu\int_{\delta-\im\infty}^{\delta+\im\infty}\frac{\d u}{u^{\nu+1}}\,\exp\left(u-\frac{z^2}{4u}\right)\,,
\ee
valid when $\text{Re}\,\nu>-1$ and $\delta>0$. We will apply this in the limit $\delta\to0^+$. Using this representation to substitute for $J_0(\Psi\sqrt{1+y^2})$ in \eqref{C1}, then rescaling $u\mapsto u\Psi^2/4$, we get
\be
    \cC = 4\pi^2\prod_{i=1}^3\int_0^\infty\d\omega_i\,\omega_i^{2\bh_i-1}\,\re^{-\eps\,\omega_i}\int_0^\infty\d y\,y\,J_0(\Phi y)\int_{\delta-\im\infty}^{\delta+\im\infty}\frac{\d u}{2\pi\im u}\,\exp\left(\frac{\Psi^2u}{4}-\frac{(1+y^2)}{u}\right)\,.
\ee
The $y$ integral is convergent when the real part of $u$ is positive, i.e., $\delta>0$. It yields
\be
    \cC = -\im\pi\prod_{i=1}^3\int_0^\infty\d\omega_i\,\omega_i^{2\bh_i-1}\,\re^{-\eps\,\omega_i}\int_{\delta-\im\infty}^{\delta+\im\infty}\d u\,\exp\left(\frac{(\Psi^2-\Phi^2)u}{4}-\frac{1}{u}\right)\,.
\ee
The factor of $\Psi^2-\Phi^2$ in the exponential may be brought to a physically interesting form:
\be\label{P2}
    \Psi^2-\Phi^2 = -4\sum_{j<k}\omega_j\omega_ks_{jk}\bar s_{jk} = -P^2\,,
\ee
where $s_{jk}$ and $\bar s_{jk}$ are as defined in \eqref{sdef}, and $P^\mu \equiv \omega_1\hat p_1^\mu+\omega_2\hat p_2^\mu+\omega_3\hat p_3^\mu$  denotes the total momentum in global coordinates.

%
%
%
\paragraph{Mellin integrals.} A standard change of variables \cite{Penedones:2016voo}  computes the Mellin integrals involved in 3-point contact Witten diagrams:
\be
\omega_i = \frac{\sqrt{t_1t_2t_3}}{t_i}
\ee
with $t_i\in(0,\infty)$ for all $i=1,2,3$. This yields 
\be
P^2=4\,t_1s_{23}\bar s_{23}+4\,t_2s_{31}\bar s_{31}+4\,t_3s_{12}\bar s_{12}\,.
\ee
To simplify the computation, we rescale the $\im\epsilon$ factors by positive factors of $\omega_i$ to send $\re^{-\epsilon\sum_j\omega_j} \mapsto \re^{-\epsilon\sum_j t_j}$; this damps the integrals at both large $\omega_i$ and large $t_i$ and will not change the $\epsilon\to 0$ limit. Doing this, we find
\be
\cC = -\frac{\im\pi}{2}\int_0^\infty\d t_1\,t_1^{\bb-1}\int_0^\infty\d t_2\,t_2^{\bc-1}\int_0^\infty\d t_3\,t_3^{\ba-1}\int_{\delta-\im\infty}^{\delta+\im\infty}\d u\,\exp\left(-\frac{P^2u}{4}-\frac{1}{u}-\eps\,{\textstyle\sum_i}t_i\right)\,,
\ee
having collected together the exponentials. The exponents $\ba,\bb,\bc$ were defined in \eqref{bardef}.

To proceed with the $t_i$ integrals, let us analyze the factor in the exponential. Notice that
\be
-\frac{P^2u}{4}-\eps\,{\textstyle\sum_i}t_i = -t_1(us_{23}\bar s_{23}+\eps)-t_2(us_{31}\bar s_{31}+\eps)-t_3(us_{12}\bar s_{12}+\eps)\,.
\ee
Since $u\in\delta+\im\,\R$, the $t_i$ integrals will converge if $\delta s_{ij}\bar s_{ij}+\eps>0$ for every $i,j$. This is ensured by sending $\delta\to0$ before we take the $\eps\to0$ limit, so this is the order of limits that we will employ. Thereafter, the Mellin integrals lead to
\be
\cC = -\frac{\im\pi}{2}\int_{\delta-\im\infty}^{\delta+\im\infty}\d u\;\re^{-1/u}\;\frac{\Gamma(\ba)\Gamma(\bb)\Gamma(\bc)}{(us_{12}\bar s_{12}+\eps)^\ba(us_{23}\bar s_{23}+\eps)^\bb(us_{31}\bar s_{31}+\eps)^\bc}\,,
\ee
where the $\delta\to0^+$ limit is implicitly understood.

Let $u=\delta+\im v$, with $v\in\R$. Then we can write
\be
    \cC = \frac{\pi}{2}\int_{-\infty}^{\infty}\d v\;\re^{\im/(v-\im\delta)}\;\frac{\Gamma(\ba)\Gamma(\bb)\Gamma(\bc)}{(\im vs_{12}\bar s_{12}+\eps)^\ba(\im vs_{23}\bar s_{23}+\eps)^\bb(\im vs_{31}\bar s_{31}+\eps)^\bc}\,.
\ee
That is, we have replaced factors like $(\im vs_{12}\bar s_{12}+\delta s_{12}\bar s_{12}+\eps)^\ba$ by $(\im vs_{12}\bar s_{12}+\eps)^\ba$ using the fact that they describe the same branch prescription for $(\im v s_{12}\bar s_{12})^\ba$ in the limit that $\delta\to0^+$ faster than $\eps\to0^+$ (because $\delta s_{12}\bar s_{12}+\eps$ stays positive).

The presence of $\delta$ in $\re^{\im/(v-\im\delta)}$ regulates the singularity of $\re^{\im/v}$ at $v=0$. We can instead set $\delta=0$ and perform the integral by a principal value prescription. Breaking the $v$ integral into the two domains $v>0$ and $v<0$ and pulling out factors of $\im |v|$ from the denominators yields
\begin{align}
    \cC &= \frac{\pi}{2}\,\frac{\Gamma(\ba)\Gamma(\bb)\Gamma(\bc)}{(s_{12}\bar s_{12}-\im\eps)^\ba(s_{23}\bar s_{23}-\im\eps)^\bb(s_{31}\bar s_{31}-\im\eps)^\bc}\int_0^\infty\d v\,\re^{\im/v}\,v^{-2-\frac\beta2}\,\re^{-\frac{\im\pi}{2}(2+\frac\beta2)}\nonumber\\
    &\hspace{2cm} + \frac{\pi}{2}\,\frac{\Gamma(\ba)\Gamma(\bb)\Gamma(\bc)}{(s_{12}\bar s_{12}+\im\eps)^\ba(s_{23}\bar s_{23}+\im\eps)^\bb(s_{31}\bar s_{31}+\im\eps)^\bc}\int_0^\infty\d v\,\re^{-\im/v}\,v^{-2-\frac\beta2}\,\re^{\frac{\im\pi}{2}(2+\frac\beta2)}
\end{align}
where we have simplified the phases by recalling that $\ba+\bb+\bc=2+\beta/2$. We have also substituted $v\mapsto-v$ for $v\in(-\infty,0)$ to obtain the second line. 

The integrals over $v$ become standard Euler integrals upon substituting $v\mapsto 1/v$. They lead to the following final result for the leaf amplitude
\be\label{Cpfin}
    \cC(\sigma_i, \bar \s_i) = \frac{\im\pi\cN/2}{(s_{12}\bar s_{12}+\im\eps)^{\ba}(s_{23}\bar s_{23}+\im\eps)^{\bb}(s_{31}\bar s_{31}+\im\eps)^{\bc}}
    -  \frac{\im\pi\cN/2}{(s_{12}\bar s_{12}-\im\eps)^{\ba}(s_{23}\bar s_{23}-\im\eps)^{\bb}(s_{31}\bar s_{31}-\im\eps)^{\bc}}\,,
\ee
with an overall normalization given by
\be\label{Ndef}
    \cN = \Gamma\bigg(1+\frac\beta2\bigg)\Gamma(\ba)\Gamma(\bb)\Gamma(\bc)\,.
\ee
We see that the $\pm \im\eps$ regulators which distinguish bulk positive and negative frequency conformal primary wave functions in \eqref{fd} also ultimately regulate the light cone singularities of the correlation functions on the celestial torus $\CT$, even though bulk time is not the the same thing as boundary time. 

The relative minus sign in \eqref{Cpfin} will be crucial for what follows. Notice that due to phase differences the two terms cancel only in the region where the $s_{jk}\bar s_{jk}$ are all positive. 

Similarly, sending $\bs_i\mapsto-\bs_i$ for all $i$ and applying the identity
\be
(-x\pm\im\eps)^\Delta = \re^{\pm\im\pi\Delta}(x\mp\im\eps)^\Delta\,,\qquad x\in\R-0\,,
\ee
we obtain the leaf amplitude associated to the spacelike wedge $W^S$ of Klein space,
\be\label{Cmfin}
    \cC(\sigma_i, -\bar \s_i) = \frac{\re^{-\im\pi\beta/2}\,\im\pi\cN/2}{(s_{12}\bar s_{12}-\im\eps)^{\ba}(s_{23}\bar s_{23}-\im\eps)^{\bb}(s_{31}\bar s_{31}-\im\eps)^{\bc}}
    -  \frac{\re^{\im\pi\beta/2}\,\im\pi\cN/2}{(s_{12}\bar s_{12}+\im\eps)^{\ba}(s_{23}\bar s_{23}+\im\eps)^{\bb}(s_{31}\bar s_{31}+\im\eps)^{\bc}}\,.
\ee


\section{Recovering PSS from leaf amplitude poles}

\label{sec:pss}

The leaf amplitudes \eqref{leafy}, \eqref{Cpfin} and \eqref{Cmfin} sum up to generate an $\im\eps$-regulated expression for the 3-gluon MHV celestial amplitude in global coordinates:
\be\label{eq:gltp}
\begin{split}
  &\cA(1^-2^-3^+) = {\delta(\beta) \over 8\pi^3}(\cL(\sigma_i, \bar \s_i)+\cL(\sigma_i, -\bar \s_i))\\&=   \frac{\cN}{8\pi^2}\,\delta(\beta) \sin\!\bigg(\frac{\pi\beta}{4}\bigg)\,\frac{s^3_{12}}{s_{23}s_{31}}\\
  &\times \bigg\{\frac{\re^{\im\pi\beta/4}}{(s_{12}\bar{s}_{12}+\im\eps)^{\ba}(s_{23}\bar{s}_{23}+\im\eps)^{\bb}(s_{31}\bar{s}_{31}+\im\eps)^{\bc}}
    + \frac{\re^{-\im\pi\beta/4}}{(s_{12}\bar{s}_{12}-\im\eps)^{\ba}(s_{23}\bar{s}_{23}-\im\eps)^{\bb}(s_{31}\bar{s}_{31}-\im\eps)^{\bc}}\bigg\}\,.
\end{split}
\ee
The normalization factor $\cN$ from  \eqref{Ndef} 
reduces to $\cN=\Gamma(\ba)\Gamma(\bb)\Gamma(\bc)$ on the support of $\delta(\beta)$. We now show that \eqref{eq:gltp} reproduces the familiar 3-gluon PSS amplitude \eqref{A3} computed by  Mellin transform from momentum space. 

The key observation is that while the gluon 3-point function in \eqref{eq:gltp} may naively appear to vanish because of the $\delta(\beta)\sin\pi\beta/4$ factor, it contains a pole in $\beta$ that cancels the zero of $\sin\pi\beta/4$. The residue at this pole gives the familiar gluon 3-point amplitude.

The first step in proving this requires reexpressing \eqref{eq:gltp} in planar coordinates $z_i,\bz_i$. This is accomplished using the conformal transformation \eqref{ltog}. Substituting
\be
\sin\s_{ij} = z_{ij}\cos\s_i\cos\s_j\,,\qquad\sin\bs_{ij} = \bz_{ij}\cos\bs_i\cos\bs_j
\ee
and multiplying \eqref{eq:gltp} by the Jacobian $\prod_{j=1}^3|\cos\s_j|^{2h_j}|\cos\bs_j|^{2\bh_j}$, we obtain an $\im\eps$-regulated version of the 3-gluon amplitude in planar coordinates,
\be\label{A3broken}
\cA(1^-2^-3^+) = \frac{\cN}{8\pi^2}\,\delta(\beta)\sin\!\bigg(\frac{\pi\beta}{4}\bigg)\,\big(\re^{\im\pi\beta/4}\cB_++\re^{-\im\pi\beta/4}\cB_-\big)\,.
\ee
The building blocks $\cB_\pm$ entering this decomposition are given by
\be
\cB_\pm = \frac{z_{12}^3}{z_{23}z_{31}}\,\frac{1}{(\veps_1\veps_2z_{12}\bz_{12}\pm\im\eps)^{\ba}(\veps_2\veps_3z_{23}\bz_{23}\pm\im\eps)^{\bb}(\veps_3\veps_1z_{31}\bz_{31}\pm\im\eps)^{\bc}}\,.
\ee
While obtaining these, we have reinstated $\veps_i = \sgn(\cos\s_i\cos\bs_i)$ by noting that
\be
\frac{\sin\s_{ij}\sin\bs_{ij}}{|\cos\s_i\cos\bs_i\cos\s_j\cos\bs_j|} = \veps_i\veps_jz_{ij}\bz_{ij}\,.
\ee
Each building block $\cB_\pm$ transforms as a 3-point Lorentzian CFT$_2$ correlator with weights $(h_i,\bh_i)$. This recasts our hyperbolic decomposition of the 3-gluon amplitude in standard notation.

\paragraph{Recovering the distributions.} Next, we show that each of the building blocks $\cB_\pm$ behaves like a contact term $\delta(\bz_{13})\delta(\bz_{23})$ near its singularity $\bz_1=\bz_2=\bz_3$, with a coefficient containing a simple pole at $\beta=0$. Again, it will suffice to prove this for $\cB_+$, the proof for $\cB_-$ being completely analogous.\footnote{We remark that by parity symmetry, each of $\cB_\pm$ will also contain the contact term $\delta(z_{13})\delta(z_{23})$ in an expansion around $z_1=z_2=z_3$ with a similar pole in $\beta$. As usual, this drops out of the MHV amplitude due to the vanishing of the Parke-Taylor prefactor $z_{12}^3/z_{23}z_{31}$. But it will become the dominant term if one studies the $\overline{\text{MHV}}$ amplitude instead (in which case one will also need to replace the $\bh_i$ by the $h_i$, etc.).}

To start the proof, set
\be\label{zetadef}
\zeta_{ij} = \sgn(\veps_i\veps_jz_{ij})\,.
\ee
We can pull out the factors of $z_{ij}$ from the denominator of $\cB_+$ to factorize it into
\begin{equation}
    \cB_+ = \frac{\sgn(z_{12}z_{23}z_{31})}{|z_{12}|^{a}|z_{23}|^{b}|z_{31}|^{c}}\,\frac{1}{(\zeta_{12}\bz_{12}+\im\eps)^\ba(\zeta_{23}\bz_{23}+\im\eps)^\bb(\zeta_{31}\bz_{31}+\im\eps)^\bc}\,.
\end{equation}
For separated insertions, $\cB_+$ manifestly contains no poles in $\beta$ and vanishes when multiplied by $\delta(\beta)\sin(\pi\beta/4)$. To show that it contains a distributional term with a pole at $\beta=0$ when all $\bz_i$ are coincident, we need to evaluate the integral
\begin{equation}
\label{eq:Iint}
    \begin{split}
        I &= \int_{-\infty}^\infty\d\bz_1\int_{-\infty}^\infty\d\bz_2\;\cB_+ \\
        &= \frac{\epsilon^{-\beta/2}\,\sgn(z_{12}z_{23}z_{31})}{|z_{12}|^{a}|z_{23}|^{b}|z_{31}|^{c}}\int_{-\infty}^\infty\d x\int_{-\infty}^\infty\d y\; \big\{\im+\zeta_{12}(x-y)\big\}^{-\bar{a}}(\im+\zeta_{23}y)^{-\bar{b}}(\im-\zeta_{31}x)^{-\bar{c}}\,,
    \end{split}
\end{equation}
where we have substituted $\bz_{1}=\bz_3+\eps x$, $\bz_{2}=\bz_3+\eps y$ to obtain the second line. Clearly we will be able to take the $\eps\to0$ limit trivially once we have taken the $\beta\to0$ limit.

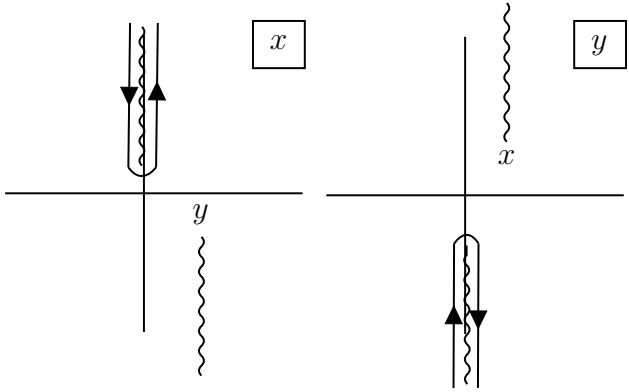
\begin{figure}[h!]
\begin{center}

\tikzset{every picture/.style={line width=0.75pt}} 

\begin{tikzpicture}[x=0.75pt,y=0.75pt,yscale=-1,xscale=1]

\draw    (86,126) -- (236,126) ;
\draw    (156,196) -- (156,46) ;
\draw    (248,127) -- (398,127) ;
\draw    (318,197) -- (318,47) ;
\draw    (155,42) .. controls (156.67,43.67) and (156.67,45.33) .. (155,47) .. controls (153.33,48.67) and (153.33,50.33) .. (155,52) .. controls (156.67,53.67) and (156.67,55.33) .. (155,57) .. controls (153.33,58.67) and (153.33,60.33) .. (155,62) .. controls (156.67,63.67) and (156.67,65.33) .. (155,67) .. controls (153.33,68.67) and (153.33,70.33) .. (155,72) .. controls (156.67,73.67) and (156.67,75.33) .. (155,77) .. controls (153.33,78.67) and (153.33,80.33) .. (155,82) .. controls (156.67,83.67) and (156.67,85.33) .. (155,87) .. controls (153.33,88.67) and (153.33,90.33) .. (155,92) .. controls (156.67,93.67) and (156.67,95.33) .. (155,97) .. controls (153.33,98.67) and (153.33,100.33) .. (155,102) .. controls (156.67,103.67) and (156.67,105.33) .. (155,107) .. controls (153.33,108.67) and (153.33,110.33) .. (155,112) -- (155,112) ;
\draw    (185,148) .. controls (186.67,149.67) and (186.67,151.33) .. (185,153) .. controls (183.33,154.67) and (183.33,156.33) .. (185,158) .. controls (186.67,159.67) and (186.67,161.33) .. (185,163) .. controls (183.33,164.67) and (183.33,166.33) .. (185,168) .. controls (186.67,169.67) and (186.67,171.33) .. (185,173) .. controls (183.33,174.67) and (183.33,176.33) .. (185,178) .. controls (186.67,179.67) and (186.67,181.33) .. (185,183) .. controls (183.33,184.67) and (183.33,186.33) .. (185,188) .. controls (186.67,189.67) and (186.67,191.33) .. (185,193) .. controls (183.33,194.67) and (183.33,196.33) .. (185,198) .. controls (186.67,199.67) and (186.67,201.33) .. (185,203) .. controls (183.33,204.67) and (183.33,206.33) .. (185,208) .. controls (186.67,209.67) and (186.67,211.33) .. (185,213) .. controls (183.33,214.67) and (183.33,216.33) .. (185,218) -- (185,218) ;
\draw    (339,30) .. controls (340.67,31.67) and (340.67,33.33) .. (339,35) .. controls (337.33,36.67) and (337.33,38.33) .. (339,40) .. controls (340.67,41.67) and (340.67,43.33) .. (339,45) .. controls (337.33,46.67) and (337.33,48.33) .. (339,50) .. controls (340.67,51.67) and (340.67,53.33) .. (339,55) .. controls (337.33,56.67) and (337.33,58.33) .. (339,60) .. controls (340.67,61.67) and (340.67,63.33) .. (339,65) .. controls (337.33,66.67) and (337.33,68.33) .. (339,70) .. controls (340.67,71.67) and (340.67,73.33) .. (339,75) .. controls (337.33,76.67) and (337.33,78.33) .. (339,80) .. controls (340.67,81.67) and (340.67,83.33) .. (339,85) .. controls (337.33,86.67) and (337.33,88.33) .. (339,90) .. controls (340.67,91.67) and (340.67,93.33) .. (339,95) .. controls (337.33,96.67) and (337.33,98.33) .. (339,100) -- (339,100) ;
\draw    (149,40) -- (148.14,112.83) ;
\draw [shift={(148.51,81.41)}, rotate = 270.68] [fill={rgb, 255:red, 0; green, 0; blue, 0 }  ][line width=0.08]  [draw opacity=0] (8.93,-4.29) -- (0,0) -- (8.93,4.29) -- cycle    ;
\draw    (163,40) -- (162.14,112.83) ;
\draw [shift={(162.65,69.91)}, rotate = 90.68] [fill={rgb, 255:red, 0; green, 0; blue, 0 }  ][line width=0.08]  [draw opacity=0] (8.93,-4.29) -- (0,0) -- (8.93,4.29) -- cycle    ;
\draw    (148.14,112.83) .. controls (155.14,122.83) and (162.14,112.83) .. (162.14,112.83) ;
\draw    (319.17,222.27) .. controls (317.49,220.62) and (317.48,218.95) .. (319.13,217.27) .. controls (320.78,215.6) and (320.77,213.93) .. (319.1,212.27) .. controls (317.43,210.61) and (317.42,208.94) .. (319.07,207.27) .. controls (320.72,205.59) and (320.71,203.92) .. (319.03,202.27) .. controls (317.36,200.61) and (317.35,198.94) .. (319,197.27) .. controls (320.65,195.6) and (320.64,193.93) .. (318.97,192.27) .. controls (317.3,190.61) and (317.29,188.94) .. (318.94,187.27) .. controls (320.59,185.59) and (320.58,183.92) .. (318.9,182.27) .. controls (317.23,180.61) and (317.22,178.94) .. (318.87,177.27) .. controls (320.52,175.6) and (320.51,173.93) .. (318.84,172.27) .. controls (317.16,170.62) and (317.15,168.95) .. (318.8,167.27) .. controls (320.45,165.6) and (320.44,163.93) .. (318.77,162.27) .. controls (317.1,160.61) and (317.09,158.94) .. (318.74,157.27) -- (318.7,152.27) -- (318.7,152.27) ;
\draw    (324.46,224.22) -- (324.74,151.39) ;
\draw [shift={(324.58,194.3)}, rotate = 270.22] [fill={rgb, 255:red, 0; green, 0; blue, 0 }  ][line width=0.08]  [draw opacity=0] (8.93,-4.29) -- (0,0) -- (8.93,4.29) -- cycle    ;
\draw    (312.14,224.33) -- (312.42,151.49) ;
\draw [shift={(312.29,182.91)}, rotate = 90.22] [fill={rgb, 255:red, 0; green, 0; blue, 0 }  ][line width=0.08]  [draw opacity=0] (8.93,-4.29) -- (0,0) -- (8.93,4.29) -- cycle    ;
\draw    (324.74,151.39) .. controls (318.51,141.44) and (312.42,151.49) .. (312.42,151.49) ;
\draw   (211,39) -- (237.14,39) -- (237.14,62.83) -- (211,62.83) -- cycle ;
\draw   (373,39) -- (399.14,39) -- (399.14,62.83) -- (373,62.83) -- cycle ;

\draw (218,45) node [anchor=north west][inner sep=0.75pt]    {$x$};
\draw (380,45) node [anchor=north west][inner sep=0.75pt]    {$y$};
\draw (179,130) node [anchor=north west][inner sep=0.75pt]    {$y$};
\draw (333,103) node [anchor=north west][inner sep=0.75pt]    {$x$};

\end{tikzpicture}
\end{center}
\caption{The contour deformation used to evaluate the integral \eqref{eq:Iint}. If the $\zeta_{ij}$ are \emph{not} all identical, both branch cuts would lie on the same side of the origin in either the $x$ plane or the $y$ plane, implying that the integral would vanish. \label{fig:contplot}}
\end{figure}

We can perform the $x,y$ integrals by contour rotation. There are two branch points in each of the $x$ and $y$ complex planes. When all the $\zeta_{ij}$ are identical, we place our branch cuts one above the real axis and one below. This is displayed in figure \ref{fig:contplot} for the case $\zeta_{12}=\zeta_{23}=\zeta_{31}=1$.

Note that if the $\zeta_{ij}$ are not all identical, the branch points will lie on the same side of the real axis in either the $x$ or the $y$ complex plane. So we can close the contour of either the $x$ or the $y$ integral and shrink it to a point. Ergo, we find 
\begin{equation}\label{Idef}
    I = \frac{\epsilon^{-\beta/2}\,\sgn(z_{12}z_{23}z_{31})}{|z_{12}|^{a}|z_{23}|^{b}|z_{31}|^{c}}\,\big\{\Theta(\zeta_{12})\Theta(\zeta_{23})\Theta(\zeta_{31})\,I_{+} + \Theta(-\zeta_{12})\Theta(-\zeta_{23})\Theta(-\zeta_{31})\,I_-\big\}
\end{equation}
where (before any contour rotations) we have defined
\begin{equation}\label{Ipm}
    I_\pm(\ba,\bb,\bc) = \int_{-\infty}^\infty\d x\int_{-\infty}^\infty\d y\;\big\{\im\pm(x-y)\big\}^{-\bar{a}}(\im\pm y)^{-\bar{b}}(\im\mp x)^{-\bar{c}}\,.
\end{equation}
Let us first look at $I_+$. Closing the $x$ and $y$ contours around the branch cuts starting at $x=\im$, $y=-\im$ leads to
\begin{equation}
\begin{split}
    I_+ &= \re^{-\im\pi\bar{a}/2}(\re^{\im\pi\bar{b}/2}-\re^{-3\im\pi\bar{b}/2})(\re^{\im\pi\bar{c}/2}-\re^{-3\im\pi\bar{c}/2})\int_0^\infty\d x\int_0^\infty\d y\;(x+y+3)^{-\bar{a}}y^{-\bar{b}}x^{-\bar{c}} \\
    &=\frac{4\pi^2\Gamma(\beta/2)}{\Gamma(\bar{a})\Gamma(\bar{b})\Gamma(\bar{c})}\;3^{-\beta/2}\,\re^{-\im\pi\beta/4}.
\end{split}
\end{equation}
Exchanging the dummy integration variables $x,y$ in \eqref{Ipm} shows that $I_-(\ba,\bb,\bc)=I_+(\ba,\bc,\bb)$. But since the integrated result for $I_+$ is totally symmetric in $\ba,\bb,\bc$, we find
\begin{equation}
    I_- = I_+\,. 
\end{equation}
As a result, the step functions in \eqref{Idef} combine to produce $\Theta(\zeta_{23}/\zeta_{12})\,\Theta(\zeta_{31}/\zeta_{12})$. Recalling \eqref{zetadef}, this is seen to be \emph{exactly} the combination of step functions present in the 3-gluon amplitude \eqref{A3}.

Both $I_+$ and $I_-$ have poles in $\beta$ with residue
\begin{equation}
    \mathrm{Res}_{\beta = 0} I_\pm = \frac{8\pi^2}{\Gamma(\bar{a})\Gamma(\bar{b})\Gamma(\bar{c})}\,.
\end{equation}
This leads to a $1/\beta$ pole in $I$ with residue
\begin{equation}
    \lim_{\beta\to 0}\beta \int \d\bz_1\,\d\bz_2\,\cB_+ = \mathrm{Res}_{\beta = 0}I = \frac{8\pi^2}{\Gamma(\bar{a})\Gamma(\bar{b})\Gamma(\bar{c})}\,\frac{\sgn(z_{12}z_{23}z_{31})}{|z_{12}|^{a}|z_{23}|^{b}|z_{31}|^{c}}\,\Theta\bigg(\frac{\veps_3z_{23}}{\veps_1z_{12}}\bigg)\,\Theta\bigg(\frac{\veps_3z_{31}}{\veps_2z_{12}}\bigg)\,.
\end{equation}
Because $\lim_{\beta\to 0}\beta\cB_+$ vanishes at generic points, we have that 
\begin{equation}\label{resBp}
    \lim_{\beta\to 0}\beta \cB_+ = \delta(\bz_{13})\,\delta(\bz_{23})\,\mathrm{Res}_{\beta = 0}I.
\end{equation}
An identical result is obtained for the other building block $\cB_-$,
\be
\lim_{\beta\to 0}\beta \cB_- = \lim_{\beta\to 0}\beta \cB_+\,,
\ee
and these can now be recombined into the full 3-gluon amplitude.

Expanding $\sin(\pi\beta/4)$ around $\beta=0$ in \eqref{A3broken} and localizing on the support of $\delta(\beta)$, we get 
\be
\cA(1^-2^-3^+) = \frac{\cN}{8\pi^2}\,\delta(\beta)\,\lim_{\eps\to0}\lim_{\beta\to0}\frac{\pi\beta}{4}\left(\cB_++\cB_-\right)\,.
\ee
The $\beta\to0$ limits of $\beta\cB_\pm$ extract the residues of $\cB_\pm$ at $\beta=0$. The latter are independent of $\eps$, so the $\eps\to0$ limit is taken trivially. Recalling the value of $\cN$ from \eqref{Ndef}, the factor of $8\pi^2/\Gamma(\ba)\Gamma(\bb)\Gamma(\bc)$ in \eqref{resBp} can be cancelled against $\cN/8\pi^2$ when $\beta=0$. Therefore, we obtain
\be
\cA(1^-2^-3^+) = \frac{\pi}{2}\,\delta(\beta)\,\frac{\sgn(z_{12}z_{23}z_{31})}{|z_{12}|^{a}|z_{23}|^{b}|z_{31}|^{c}}\,\delta(\bz_{13})\,\delta(\bz_{23})\,\Theta\bigg(\frac{\veps_3z_{23}}{\veps_1z_{12}}\bigg)\,\Theta\bigg(\frac{\veps_3z_{31}}{\veps_2z_{12}}\bigg)
\ee
which agrees with the PSS 3-gluon amplitude \eqref{A3}, concluding the proof.






\section*{Acknowledgements}

We are grateful to Mina Himwich for many useful conversations. This work was supported by DOE grant de-sc/0007870, NSF GRFP grant DGE1745303, the Simons Collaboration on Celestial Holography and the  Gordon and Betty Moore Foundation and the John Templeton Foundation via the Black Hole Initiative.

\bibliographystyle{JHEP}
\bibliography{refs}

\end{document}